\documentclass[conference]{IEEEtran}
\IEEEoverridecommandlockouts
\usepackage{cite}
\usepackage{amsmath,amssymb,amsfonts}
\usepackage{algorithm}
\usepackage{algorithmic}
\usepackage{graphicx}
\usepackage{textcomp}
\usepackage{xcolor}
\usepackage{multirow}
\usepackage{subfigure}

\def\BibTeX{{\rm B\kern-.05em{\sc i\kern-.025em b}\kern-.08em
    T\kern-.1667em\lower.7ex\hbox{E}\kern-.125emX}}
\begin{document}

\title{Hierarchical Reinforcement Learning Based Video Semantic Coding for Segmentation}
%\title{Semantic coding for video segmentation}
% \title{Hierarchical Reinforcement Learning Based task-driven Video Semantic Coding}

%\author{Guangqi Xie, Xin Li, Shiqi Lin, Zhibo Chen}

\author{\IEEEauthorblockN{1\textsuperscript{nd} Guangqi Xie, Xin Li, Shiqi Lin, Zhibo Chen\textsuperscript{*}\thanks{\textsuperscript{*}Corresponding author}}
\IEEEauthorblockA{\textit{Unviersity of Science and Technology of China, Hefei, China} \\
\{jszjxgq, lixin666, linsq047\}@mail.ustc.edu.cn, \\ chenzhibo@ustc.edu.cn}
\and \IEEEauthorblockN{2\textsuperscript{nd} Li Zhang, Kai Zhang, Yue Li}
 \IEEEauthorblockA{\textit{ByteDance Inc.} \\
\{lizhang.idm, yue.li\}@bytedance.com \\
kzhang1981@hotmail.com
}}
% \and
% \and
% \and
% \and
\newcommand{\ieno}{\textit{i.e.}}
\newcommand{\egno}{\textit{e.g.}}

\maketitle

\begin{abstract}
The rapid development of intelligent tasks, \egno, segmentation, detection, and classification, etc, has brought an urgent need for semantic compression, which aims to reduce the compression cost while maintaining the original semantic information. However, it is impractical to directly integrate the semantic metric into the traditional codecs since they cannot be optimized in an end-to-end manner. To solve this problem, some pioneering works have applied reinforcement learning to implement image-wise semantic compression. Nevertheless, the video semantic compression has not been explored since its complex reference architectures and compression modes. In this paper, we take a step forward to video semantic compression and propose the \textbf{H}ierarchical \textbf{R}einforcement \textbf{L}earning based task-driven \textbf{V}ideo \textbf{S}emantic \textbf{C}oding, named as HRLVSC. Specifically, to simplify the complex mode decision of video semantic coding, we divided the action space into frame-level and CTU-level spaces in a hierarchical manner, and then explore the best mode selection for them progressively with the cooperation of frame-level and CTU-level agents. Moreover, since the modes of video semantic coding will exponentially increase with the number of frames in a Group of Pictures (GOP), we carefully investigate the effects of different mode selections for video semantic coding, and design a simple but effective mode simplification strategy for it. We have validated our HRLVSC on video segmentation task with HEVC reference software HM16.19. Extensive experimental results demonstrated that our HRLVSC can achieve over 39\% BD-rate saving for video semantic coding under the Low Delay P configuration.

\end{abstract}

\begin{IEEEkeywords}
%semantic video coding, task-driven bit allocation, rate-distortion optimization, hierarchical reinforcement learning
semantic video coding for segmentation, rate-distortion optimization, hierarchical reinforcement learning
\end{IEEEkeywords}

\section{Introduction}
Deep learning has brought a technological revolution in visual intelligent tasks, such as image/video object detection~\cite{VOD1,VOD7,VOD3,VOD6}, segmentation~\cite{MiVOS,VOS1,VOS3,VOS4}, face detection~\cite{FDSurvey1,FDBenchmark}, and person reidentification~\cite{reid1,reid3,reid5}, etc, which enables the intelligent tasks to be greatly developed and broadly applied in people's lives. Meanwhile, billions of images/videos for intelligent tasks have caused a heavy burden for transmission and storage. However, the commonly-used image/video codecs, such as JPEG, AVC~\cite{AVC}, HEVC~\cite{HEVC},  VVC~\cite{VVC}, and related techniques~\cite{li2020multi,li2022hst} are usually designed and optimized with pixel-level metrics, such as PSNR~\cite{PSNR}, and perceptual metrics~\cite{liu2022swiniqa,zhang2018unreasonablelpips}, which are not optimal for semantic compression. It is crucial to investigate how to reduce the compression cost while maintaining the semantic information of images/videos according to specific intelligent tasks (\ieno, task-driven semantic coding).

Different from the pixel-wise fidelity (\egno, PSNR ) and perceptual fidelity (\egno,  MS-SSIM), task-driven semantic fidelity is hard to be integrated into traditional codecs directly. The reason lies in that traditional codecs cannot be optimized in an end-to-end manner such as learning-based coding(\egno, LBHIC\cite{LBHIC}). Furthermore, traditional codecs conduct image/video compression on pixel level, while semantic information cannot be represented with simple pixel-wise metrics like PSNR. To tackle the above challenges, some works~\cite{TaskDriven,RLBA} take a step forward to task-driven semantic coding, and then introduce the reinforcement learning to implement the Rate-Distortion Optimization (RDO) of semantic coding. Despite that these studies have achieved great performances on many image intelligent tasks, (including image segmentation, detection and classification), their methods are only designed for the all-intra mode of semantic coding instead of unified video semantic coding. Therefore, task-driven video semantic coding is still under-explored.

%Notwithstanding, traditional video coding standards, such as H.264/AVC\cite{AVC}, H.265/HEVC\cite{HEVC}, H.266/VVC\cite{VVC}, can hardly be optimized in such an end-to-end manner. They are mostly established on the block-based hybrid coding framework which are usually optimized block by block, module by module. As a result, the semantic metrics can not be integrated into traditional hybrid coding framework seamlessly. 

%To solve the aforementioned issues, work\cite{RLBA, task-driven} firstly introduced a semantic importance map to measure the task-related pixel-wise semantic fidelity, and then adopted reinforcement learning to explore the best bit allocation scheme that maximize semantic fidelity, finally implemented adaptive semantic bit allocation under traditional hybrid coding framework. However, due to the complex reference relationship and compression mode, the bit allocation algorithm remains in the spatial domain.

\begin{figure*}
\centering 
\includegraphics[width=0.8\textwidth]{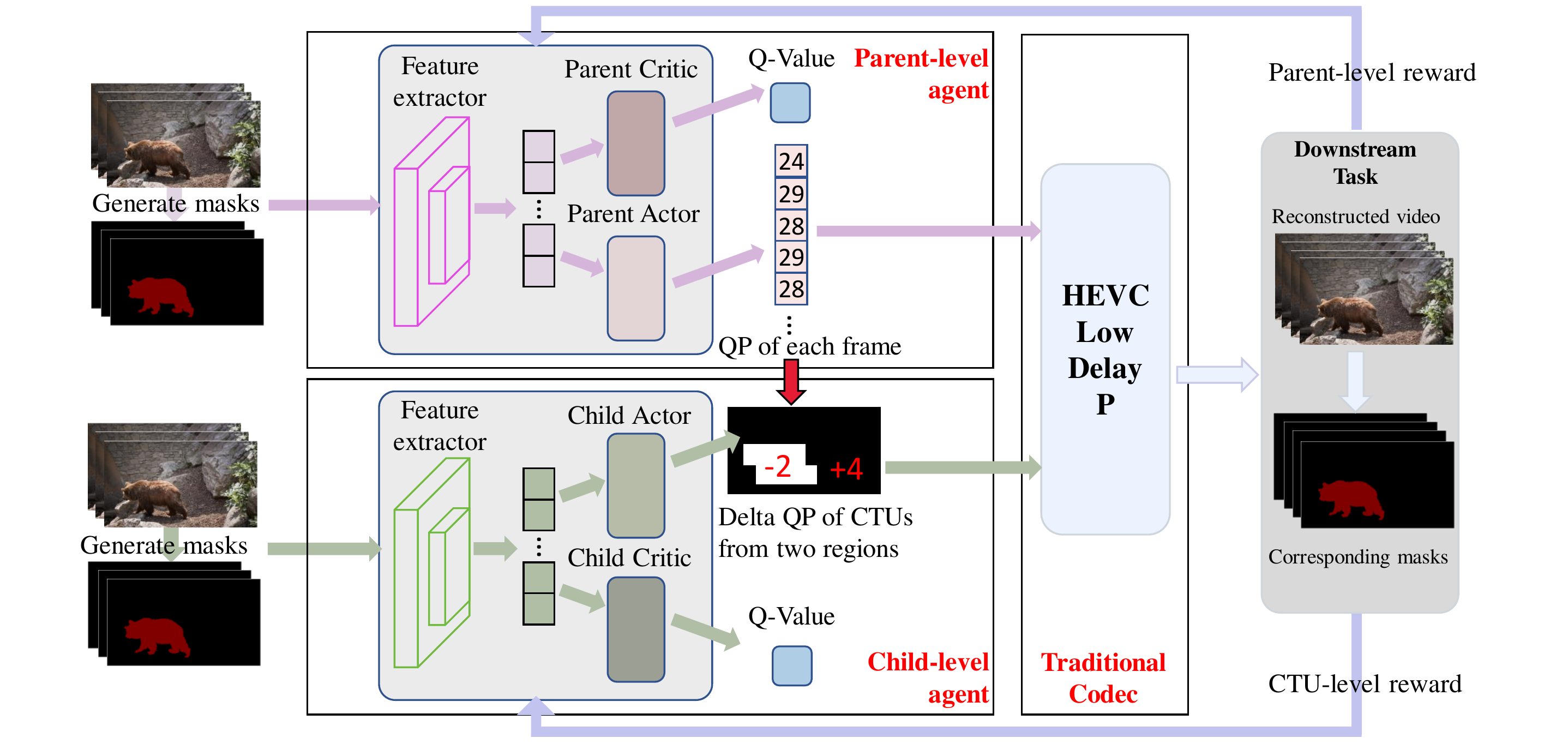}

% \includegraphics[width=0.9\textwidth]{VCIP_v2.pdf}
% \subfigure[]{\includegraphics[width=0.6\textwidth]{HRLcoding.pdf}}
% \subfigure[]{\includegraphics[width=0.3\textwidth]{network2.pdf}}
% \includegraphics[width=0.9\textwidth]{HRLVSC1.pdf}
% \includegraphics[width=0.5\textwidth]{HRLcoding.pdf}  
% \includegraphics[width=0.4\textwidth]{Network.pdf}
\caption{Illustration of our proposed HRLVSC. The hierarchically structured network generates optimal task-driven compression modes from parent level to child level progressively. In the first stage, the parent-level agent takes three consecutive frames and corresponding masks as input, extracts deep features and then predicts the quantization parameters(QP) for each frame. Finer control is at the next stage. The child-level agent extracts corresponding features and outputs the relative QP around the central frame-level QP for each CTU. Finally, traditional codec takes selected modes from both levels and conducts the optimal task-driven compression. The actual rate and semantic fidelity of each level are used as rewards to train the agents in a coarse-to-fine manner.}
\label{fig:Network}
%\vspace{-3mm}
\vspace{-5mm}
\end{figure*}

Compared with image coding, where the compression of different coding tree units (CTUs) are relatively independent\cite{CNNIntraRC}, video compression contains amounts of compression modes and reference architectures. It is complex and time-consuming to find the optimal mode selection for task-driven video semantic coding. To simplify the mode decision for video compression, previous  works~\cite{LambdaAdaptiveBA,VVCRCwithCauchy,AdaptiveQP,AdaptiveLambda,ConvexOpt,SkipbasedRC,TemporalRDO,lstm1,DRLforRCofDynamicVideo,IntraRCwithRL,FramelevelOBAwithRL} have investigated the effects of the coding modes of current frame on the following frames in one GOP, and then, model the distortion or rate propagation with linear function\cite{LambdaAdaptiveBA,VVCRCwithCauchy,AdaptiveQP,AdaptiveLambda,ConvexOpt,SkipbasedRC,TemporalRDO}, deep learning model\cite{lstm1} or reinforcement learning agent\cite{DRLforRCofDynamicVideo,IntraRCwithRL,FramelevelOBAwithRL}. 
However, the above methods only explored the relationship between different mode selections and pixel-wise fidelity, which is not suitable for task-driven semantic fidelity.

In this paper, we take a step forward to task-driven video semantic coding. 
% There are two typical chanllege for xxx.
To integrate the video semantic fidelity to the rate-distortion optimization of traditional codecs and solve the challenges brought by tremendous amounts of mode selections, we propose the hierarchical reinforcement learning based task-driven video semantic coding, named as HRLVSC. Specifically, we divide the mode selections of video compression into frame level and CTU level in a hierarchical manner, and then find the best mode selection for task-driven video semantic coding progressively with the cooperation of frame-level and CTU-level agents. Moreover, to simplify the mode decision for video semantic coding, we also carefully investigate the effects of different compression modes for task-driven video semantic coding. Based on our exploration, we design an efficient but effective mode simplification strategy for video semantic coding. To validate the effectiveness of our HRLVSC, we select video segmentation as target task. Extensive experiments under the Low Delay P configuration have demonstrated that our HRLVSC can achieve over 39\% BD-rate saving for video semantic coding.

%Since the traditional codecs cannot be optimized in an end-to-end manner together with intelligent tasks, we follow the     
%In this paper, we present a novel method for semantic bit allocation in both temporal and spatial domain. To do so, we adopted a hierarchical structure to mimicking the traditional rate control procedure, specifically, dividing the mode decision space into frame-level and CTU-level. Next, we utilize hierarchical structured agents, namely, a frame-level agent and a CTU-level agent, to explore the best mode in a coarse-to-fine manner. Moreover, considering the complex reference relationship which makes the modes of video semantic coding exponentially increase with the number of frames in a Group of Pictures (GOP), we conducted experiments to demonstrate the effects of different mode selections for video semantic coding, and design a simple but effective mode simplification strategy for it. Finally, we conducted thorough experiments to verify the effectiveness of our work.

The main contributions of our work can be summarized as follows:
\begin{itemize}
    \item  As the pioneering work, we propose the hierarchical reinforcement learning based video semantic coding(\ieno, HRLVSC) for segmentation, where the complex mode space is simplified into frame-level and CTU-level, and the optimal mode selection is learned with the cooperation of frame- and CTU-level agents in a progressive manner.
    \item We carefully explore the correlation between different mode selections and the semantic fidelity, and propose an efficient but effective mode simplification strategy for task-driven video semantic coding.
    %Hierarchical reinforcement learning: We introduce the hierarchical reinforcement learning into video semantic coding and propose a simple yet efficient solution.
    \item Extensive experiments on video segmentation task under low-delay P configuration have demonstrated the superiority of our proposed HRLVSC, which exceeds the standard software HM16.19\footnote{Available:https://hevc.hhi.fraunhofer.de/svn/svn\_HEVCSoftware/tags/HM-16.19/} by a BD-rate saving of 39\%.
    
    %Mode simplify Strategy: We propose a simple yet effective mode simplify strategy, and we achieve more than 38\% BD-rate savings based on that strategy. 
\end{itemize}

The rest of the paper is organized as follows. In sec.~\ref{sec: method}, we clarify our task-driven video semantic coding scheme HRLVSC in detail. Sec.~\ref{sec: experiments} describes our experimental setting and validates the effectiveness of our proposed HRLVSC by comparing it with the state-of-the-art codecs and a series of ablation studies. 
%introduces our task-driven video semantic coding dataset along with experiments and ablation studies. 
Finally, we conclude this paper in Section~\ref{sec: conclusion}.

\section{Proposed Method}
\label{sec: method}
In this section, we will introduce our hierarchical reinforcement learning based video semantic coding scheme (\ieno, HRLVSC) from the perspective of problem formulation, technique details and mode simplification.
%In this section, we introduce our Hierarchical Reinforcement Learning Based Semantic Video Coding(HRLSVC) scheme in the order of problem formulation, experiments analysis and concrete solution. 

\subsection{Problem formulation}
Task-driven video semantic coding aims to reduce the computation cost while maintaining the semantic information existed in videos, which can be formulated as:
\begin{equation}
    min~J_s,J_s = D(\mathcal{M})+\lambda_s \sum_{t=1}^{T_f}\sum_{i=1}^{N} R_{t,i}(\mathcal{M}),
    \label{eq:rdo}
\end{equation}
%The semantic bit allocation problem can be formulated as follows:
where $J_s$, $D(\mathcal{M})$ and  $R_{t,i}(\mathcal{M})$ are  rate-distortion performance, the semantic distortion of whole video and the rate of the $i^{th}$ CTU in the $t^{th}$ frame, respectively. $\mathcal{M}$ represents the selected mode for compression, and $\lambda_s$ is the hyperparameter to adjust the importance of rate and distortion. $T_f$ and $N$ are the number of frames and CTUs in one frame, respectively. To find the best mode $\mathcal{M^*}$ for Eq.~\ref{eq:rdo}, a straightforward method is to utilize a reinforcement learning agent to explore the optimal mode adaptively like the work~\cite{TaskDriven}. However, the optional modes for video semantic compression will exponentially increase with the number of frames and CTUs, which inevitably prevents the learning of RL agent. To simplify the exploration space and enables the RL agent to learn the optimal mode effectively and efficiently, we divide the exploration space into frame level and CTU level, and then, introduce the hierarchical reinforcement learning to solve the Eq.~\ref{eq:rdo}. Specifically, as shown in Eq.~\ref{eq:hrl}, we set two-step goals respectively for parent-level agent and child-level agent. The parent-level agent $RL_p$ aims to explore the best frame-level mode ${\mathcal{M}_f}^*$ in pursuit of minimizing the frame-level rate-distortion $J_{sf}$. And the child-level agent $RL_c$ is devoted to minimizing the final semantic rate-distortion performance $J_s$ by finding the best CTU-level mode ${\mathcal{M}_c}^*$ while cooperating with parent-level agent.

\begin{equation}
    \centering
    \begin{aligned}
    J_{sf}&=D(\mathcal{M}_{f})+\lambda_s \sum_{t=1}^{T_f} R_{t}(\mathcal{M}_{f})\\
    J_{s}&= D(\mathcal{M}_{c}|\mathcal{M}_{f}^{*})+\lambda_s \sum_{t=1}^{T_f}\sum_{i=1}^{N} R_{t,i}(\mathcal{M}_{c}|\mathcal{M}_{f}^{*}) \\
    \mathcal{M}_{f}^{*}&=RL_{p}(min~J_{sf}), \mathcal{M}_{c}^{*}=RL_c(min ~J_{s})
    \end{aligned}
    \label{eq:hrl}
\end{equation}
With hierarchical reinforcement learning in Eq.~\ref{eq:hrl}, we can obtain the best mode pair $\{{\mathcal{M}_f}^*, {\mathcal{M}_c}^*\}$ for task-driven semantic coding effectively and efficiently. In this paper, the frame-level mode $\mathcal{M}_f$ and CTU-level mode $\mathcal{M}_c$ are the Quantization Parameter (QP) of one frame and the relative QP (\ieno, $\Delta QP$) around the central frame-level QP for each CTU.  

\subsection{Hierarchical Reinforcement Learning Based Video Semantic Coding for Segmentation}

We aim to utilize the parent-level agent and child-level agent to explore the optimal frame-level mode and CTU-level mode based on the former, respectively. Therefore, we model this decision-making problem as a hierarchical MDP process that aligns well with the nature of hierarchical reinforcement learning (HRL).

A one-level RL agent commonly models the policy learning problem as a Markov decision process (MDP) represented with $({\mathcal{S}},{\mathcal{A}},{\mathcal{P}},R,\gamma,T )$. The RL agent observes the environment state $s \in {\mathcal{S}}$ and relies on the learnable policy $\pi (a\left| {s)} \right.:{\mathcal{S}}\times {\mathcal{A}} \to [0,1]$ to take an action $a \in {\mathcal{A}}$. Then, the RL agent receives a step-wise reward $r:{\mathcal{S}} \times {\mathcal{A}} \to \mathbb{R}$. 
The environment moves to next state with a transition function denoted as ${\mathcal{P}}:{\mathcal{S}} \times {\mathcal{A}} \times {\mathcal{S}} \to [0,1]$.
$\gamma \in (0,1]$ is a discount factor and $T$ is a time horizon.
% according to transition function $P:S \times A \times S \to [0,1]$.
We aim to learn an optimal policy ${\pi ^*}$ which can maximize the accumulative reward $R$.

Furthermore, HRL contains two-level RL agents, \ieno, the parent-level (frame-level) agent and the child-level (CTU-level) agent, so as to learn a parent policy ${\pi ^P}({a^P}\left| {{s^P}} \right.)$ and a child policy ${\pi ^C}({a^C}\left| {{s^C},{a^P}} \right.)$. %corresponding to $MD{P^P}\!=\!({\mathcal{S}^P},{\mathcal{A}^P},{{\mathcal{P}^P}},{R^P},{\gamma},T)$ and $MD{P^C}\!=\!({{\mathcal{S}^C}},{{\mathcal{A}^C}},{{\mathcal{P}^C}},{R^C},{\gamma},T)$, respectively. 
The parent policy outputs a parent action $a^P \in {{\mathcal{A}^P}}$, which is taken as the condition for the following decision by the child policy. Thanks to the parent-level agent which first makes decisions at frame-level, the action space of child-level agent can be effectively reduced. With a smaller action space, it's easier for the CTU-level policy to find a more effective strategy.

Here, we give the detailed design of HRL for task-driven video semantic coding. \textbf{\textit{State:}} Both the parent-level policy and the child-level policy need to perceive the content of the frames. Therefore, given three consecutive frames and corresponding masks, we concatenate them and utilize the deep features extracted by convolution network as the parent-level state and the child-level state. \textbf{\textit{Action:}} The parent-level agent is responsible for assigning {$QP$} for a frame in a coarse way. $QP$ is a central value within a range. Based on the decision of the parent-level agent, the child-level agent further determines $\triangle QP$ around the central value for each CTU in the frame. \textbf{\textit{Reward:}} The parent-level agent and the child-level agent target at minimizing the impact of compression on the performance of downstream network. We define the reward functions of parent-level and child-level policies as: 
% To be consistent with the objective of the two
%这里~！！！！！在这里写
\begin{equation}
\centering
\begin{split}
    Rw_f&=M_s\left(\mathcal{M}_{f}\right)-\lambda_s \sum_{t=1}^{T_f} R_t\left(\mathcal{M}_{f}\right)-\alpha_f\\
    Rw_c&=M_s\left(\mathcal{M}_{c} \mid P_{f}^{*}\right)-\lambda_s \sum_{t=1}^{T_f}\sum_{i=1}^{N} R_{t,i}\left(\mathcal{M}_{c} \mid \mathcal{M}_{f}^{*}\right)-\alpha_c
\end{split}
\end{equation}
where the frame-level reward $Rw_f$ is expressed as the sum of two terms: the task-related fidelity $M_s(\mathcal{M}_f)$, which is measured by the Mean Intersection over Union (mIOU) for video segmentation, and the negative sum of rate $R_t(\mathcal{M}_f)$ of each frame. $\alpha_f$ and $\alpha_c$ are hyperparameters to keep the initial reward close to zero, and then stimulus the agent to explore more optimal actions. $\lambda_s$ is an adjustable semantic coding parameter that balances the semantic distortion against rate. The CTU-level reward $Rw_c$ is similar to the frame-level reward except for the following difference. First, the distortion term is estimated by the task accuracy $M_s$ after all of the CTU-level parameters are selected, and the rate term is measured by the negative sum of the bitrate $R_{t, i}$ of all CTUs. Second, considering that the CTU-level action $\mathcal{M}_c$ is restricted by frame-level action $\mathcal{M}_f$, the CTU-level reward is consequently conditioned on $\mathcal{M}_f$. The two reward functions are consistent with the objective of the hierarchical policy in Eq.~\ref{eq:hrl}. 
\begin{figure}[t]
    \centering
    \includegraphics[width=0.9\linewidth]{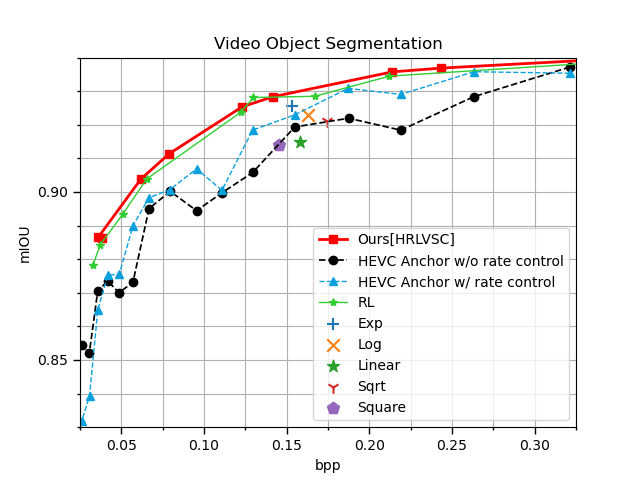}
    \caption{Comparison of proposed HRLVSC(red) against HEVC anchor without rate control(black), HEVC anchor with rate control(blue), RL(green) and hand-crafted scheme(scatter).}
    \label{fig:Ablation study}
    \vspace{-4mm}
\end{figure}
%这里也写清楚哈~我先写下面ok~~

Given that the actions/modes $QP$ and $\triangle QP$ subject to discrete distribution, therefore, we adopt the widely used Advantage Actor-Critic (A2C) algorithm \cite{mnih2016asynchronous} where the actor network aims to learn a discrete control policy while the critic network focuses on estimating the value of state $V_{\varphi}^{\pi \theta}\left(s\right)$. The parent-level and child-level agents are detailed in Fig.~\ref{fig:Network}.

\subsection{Mode simplification}
Although the HRL has reduced the action space greatly in some content, the optional modes for video semantic coding are still severely unaffordable for training since the expensive coding time cost. To further cut down the training cost, we aim to find an effective but efficient mode simplification strategy for task-driven video semantic coding. Specifically, we design the mode simplification strategy from two perspectives, \ieno, frame-level simplification and CTU-level simplification.

For frame-level actions, deciding one specific QP value for each frame in one GOP is impractical. The complexity will increase exponentially with the number of frames in one GOP.  To reduce the complexity while keeping the characteristic of original traditional codecs, we simplify the action space by only deciding the QPs for the first two frames in one GOP. The QPs of other frames in this GOP are set with the offset used in traditional codecs. For CTU-level action space, we simplify the action space based on the characteristic of semantic task, \ieno, the accuracy of semantic task are mainly associated with the semantic-related region in one frame. Therefore, we utilize the semantic mask, which is generated with corresponding task, to divide the region of one frame into two parts, \ieno, semantic-related region and semantic-unrelated region. Then, we can allocate two CTU-level QPs respectively for these two regions, without requiring to decide one QP value for each CTU.

\section{Experiments}
\label{sec: experiments}

\subsection{Dataset and Implementation Details}
To validate the effectiveness of our proposed HRLVSC, we conduct experiments on video segmentation task under the commonly-used low-delay P configuration with the reference software HM 16.19.

\noindent\textbf{Dataset:} We construct a video semantic coding dataset to optimize and validate our HRLVSC framework, named as TVSC dataset. The dataset contains one video semantic task, \ieno, video segmentation task. Our dataset is based on the commonly-used dataset DAVIS2017~\cite{davis2017} for video segmentation task. For each video, we resized them as  $960\times544$, and compressed them with the modes used in our optional action spaces. 

\noindent\textbf{Implementation Details:}
The HRL model is implemented with Pytorch platform. We train our HRL model with one NVIDIA 1080Ti GPU for 10000 iterations. The batchsize is 30 and the learning rate is 1e-3 for parent-level agent and 1e-4 for child-level agent.

\subsection{Performance Analysis}
%\tcp{In this section, we first compare our HRLVSC scheme with HEVC standard software HM 16.19 with two modes \ieno, with rate control and without rate control. Then we conduct ablation studies by replacing our HRL framework with naive RL framework used in~\cite{TaskDriven}. Finally, we compare our HRLVSC with hand-craft algorithms and make a complexity analysis.}
%To demonstrate the effectiveness of our HRLVSC, we compare our algorithm with HEVC anchor without Rate Control and HEVC anchor with Rate Control. Afterwards, 
%we conduct ablation study on the hierarchical structure against single-layer structure. Finally, we demonstrate the performance against handcrafted QP map.
\subsubsection{Compared with HEVC Anchor}
To verify the effectiveness of our HRLVSC scheme, we compare our HRLVSC with standard HEVC codec HM 16.19 under Low-delay P configuration.  
For HEVC, we select the QP from 22 to 37 for compression as our baseline. For our proposed HRLVSC, we set $\lambda$ as 0, 0.05, 0.1, 0.15, 0.2, 0.4, 0.6, 0.8, 1.0, respectively.
%We compare our algorithm with HEVC anchor with and without rate control. 
The result is shown in Table~\ref{tab:HRLvsAnchor}. From the table, we can observe that our proposed HRLVSC scheme outperforms HEVC anchor by a bit saving of 39.93\%. Even compared with HEVC anchor with rate control, our HRLVSC can still achieve a bit rate saving of 20.47\%, which demonstrates the effectiveness of our HRLVSC for task-driven video semantic coding.

% \begin{figure}[H]
%     \centering
%     \includegraphics[width=0.5\textwidth]{Demo_HRL_HM_HMRC.png}
%     \caption{Comparisons between proposed HRLVSC and HEVC anchor with or without rate control.}
%     \label{fig:HRLvsAnchor}
% \end{figure}

%AS shown in \ref{fig:HRLvsAnchor}, the proposed HRLVSC can reserve more semantic fidelity on segmentation task than HEVC. To quantify the result, we calculate the differ in the area under curve, namely, AUC-Rate and AUC-PSNR as shown in 

\begin{table}[H]
    \centering
    \vspace{-3mm}
    %\caption{BD-Rate and BD-mIOU relative to the baseline HM16.19 with and without rate control}
    \caption{BD-Rate and BD-mIOU compared with Anchor HM16.19 with fixed QP}
    \begin{tabular}{c|c|c|c}
        \hline
        Method & HEVC w/ rate control & Proposed HRLVSC & RL\\
        \hline
        BD-Rate & -19.46\% & -39.93\% & -35.54\%\\
        \hline
        BD-mIOU & 0.67\%  & 1.50\% &  1.31\%\\
        \hline
    \end{tabular}
    \label{tab:HRLvsAnchor}
    \vspace{-3mm}
\end{table}

\subsubsection{Ablation Study}
We conduct the ablation studies from two perspectives: 1) Replacing the hierarchical reinforcement learning agents with one reinforcement learning agent used in ~\cite{TaskDriven}. 2) Substituting our HRLVSC with hand-crafted methods \ieno, assigning the higher QP value for semantic-unrelated CTUs and lower QP value for semantic-related CTUs. The experimental results are as follows:

\textbf{Ablation on HRL}: As shown in~\ref{fig:Ablation study}, the performance of  applying only one reinforcement learning agent \ieno, RL in Fig.~\ref{fig:Ablation study}, for both frame-level and CTU-level QPs decision, is lower than employing hierarchical reinforcement learning.
%As shown in \ref{fig:HRLvsRL}, our hierarchical structured agents performer better than single-layer agent, 
The reasons for that are: 1) The hierarchical structure restricts the child-level action space, which makes the training procedure more stable. 2) The action spaces are complex, which is hard for RL to make decision.
%The high level agent shares a global view of the overall rate distortion performance, which will prevent the agent converges to some local optimal point.
\begin{table}[H]
\vspace{-3mm}
    \caption{Time complexity}
    \centering
    \begin{tabular}{c|c|c|c}
        \hline
        Run time & \multicolumn{2}{|c|}{Encoder} & Decoder \\
        \hline
        \multirow{2}*{Proposed} & HRL agent & 0.17s & \multirow{2}*{1.23s}\\
        \cline{2-3}
        ~&coding& 324.25s &~\\
        \hline
        % HEVC w/ Rate Control&\multicolumn{2}{|c|}{10.41s}&?s\\
        % \hline
        HEVC w/o Rate Control&\multicolumn{2}{|c|}{318.45s}&1.23s\\
        \hline
    \end{tabular}
    \label{tab:Complexity}
    \vspace{-3mm}
\end{table}
  
 \textbf{Comparison with hand-crafted schemes} In this part, we will discuss the overwhelming advantage of proposed HRL against hand-crafted schemes. 
For video segmentation task, we can obtain the segmentation masks for frames in a video. Therefore, the hand-crafted methods can assign different QP values for each CTU based on the semantic importance \ieno, the segmentation mask ratio $S$ for each CTU. In other words, when the mask ratio is higher, the CTU will be assigned lower QP value for better coding quality. For frame-level QP \ieno, $QP_f$, we keep the original QP value in standard software HM 16.19 for hand-crafted scheme. For CTU-level QP value in hand-crafted scheme, we attempt to adopt different functions to establish the relationship between QP value and semantic importance \ieno, mask ration $S$, respectively as linear, exponential, square, log, and square root functions.
    % \begin{equation}
    % \centering
    %     \begin{aligned}
    %     Linear:&QP=QP_f-10*S\\
    %     Exp:&QP=QP_f-10*(2^S-1)\\
    %     Sqrt:&QP=QP_f-10*\sqrt{S}\\
    %     Square:&QP=QP_f-10*S^2\\
    %     Log:&QP=QP_f-10*\log_2 (S+1)
    %     \end{aligned}
    % \label{eq:handcraft}
    % \end{equation}
The experimental results are shown in Fig~\ref{fig:Ablation study}, we can find that hand-crafted schemes are far from our proposed HRLVSC, since our scheme can capture the optimal relationship between QP value and semantic importance. 

\subsubsection{Complexity Analysis}
In this section, we compare our HRLVSC with standard software HM 16.19 from the perspective of time complexity. As shown in Table~\ref{tab:Complexity}, our HRLVSC does not increase any decoding time. For encoding time, our HRL agents only take about 1.52 seconds for the QP decision of the whole video with the size of 960x544.
%In this section, we exhibit the complexity of our algorithm. As shown in \ref{tab:Complexity}, our algorithm do not impact the decoder time. 
%What's more, the encoding time is almost the ? as HEVC. The QP decision takes approximately 1.52s for the whole video of size 960x544 when running with NVIDIA3080Ti. 
It is efficient and effective to apply our algorithm in the task-driven video semantic coding.

\section{Conclusion}
In this paper, we are the first to investigate the task-driven video semantic coding, and propose the hierarchical reinforcement learning based scheme HRLVSC for it. Unlike the image semantic coding, task-driven video semantic coding contains tremendous reference architectures and coding modes, which inevitably prevents its development. To tackle this challenge, we divide the complex coding modes into frame level and CTU level, and then, introduce the hierarchical RL agents for them. To further reduce the time complexity for training, we carefully design a simple but effective mode simplification strategy for task-driven video semantic coding. Extensive experiments on video segmentation task under low-delay P configuration have validated the effectiveness of our scheme. We will extend our HRLVSC to more video semantic tasks and more configuration of video coding in future work.
\label{sec: conclusion}
\bibliographystyle{IEEEtran}
\bibliography{reference}
\end{document}